\documentclass[a4paper,10pt,twocolumn,showpacs,preprintnumbers,amsmath,amssymb]{revtex4}
\usepackage{graphicx}
\usepackage{dcolumn}
\usepackage{bm}
\begin{document}
\preprint{}
\title{Nano-structured thin films growth in stochastic plasma-condensate systems}
\author{Vasyl O.~Kharchenko}\email{vasiliy@ipfcentr.sumy.ua} 
\affiliation{Institute of Applied Physics, National Academy of Sciences of
Ukraine, 58 Petropavlovskaya St., 40000 Sumy, Ukraine}
\affiliation{Sumy State University, 2 Rimskii-Korsakov St., 40007 Sumy, Ukraine}
\author{Alina V. Dvornichenko}
\affiliation{Sumy State University, 2 Rimskii-Korsakov St., 40007 Sumy, Ukraine}
\begin{abstract}
We derive the stochastic model of plasma-condensate systems by taking into account anisotropy 
in transference of adatoms between neighbor layers and by introducing fluctuations of adsorbate 
flux. We show, that by varying the fluctuation's intensity on can govern dynamics of pattern formation 
on intermediate layer of multi-layer plasma-condensate system. It is shown that the morphology of 
the growing surface, type of surface structures and their linear size can be controlled by the intensity 
of the adsorbate flux fluctuations.     
\end{abstract}
\maketitle


\section{Introduction}

Nano-structured thin films attracts an increasable attention in last few decades due to their 
frequently usage in modern micro- and nano-electronic devices. There are several techniques 
allowing one to produce thin films with well structures surface. Among them one can issue 
epitaxy and deposition from gaseous phase or plasma. During epitaxial growth pyramidal structures 
are realized on a growing surface \cite{MBE0,MBE1,PhysScr11}. In gas-condensate systems depending 
on the deposition conditions one can produce different types of patterns: periodic array of vacancy 
islands \cite{cite25}, elongated islands of adsorbed semiconductors \cite{cite26,cite27} and metals 
\cite{cite28} and spherical vacancy islands (nano-holes) or islands of adsorbate (nano-dots) 
\cite{cite33,cite34,cite35,cite36,cite37}. Plasma-condensate devices are widely used to produce 
separated spherical adsorbate structures on the surface with small linear size \cite{Perekrestov1,Perekrestov2}. 
In such systems the growing surface is located in a hollow cathode \cite{Per8}. 
Under the influence of the electric field near substrate a part of adatoms can desorb back to plasma and after 
additional ionization adsorb on the upper layers of multi-layer growing surface \cite{Per8}. 
These processes lead to the bias in the vertical diffusion of adatoms between layers with preferential motion 
from lower layers towards upper ones and the strength of the electric field defines the anisotropy strength of 
these transitions.  

With the development of computer technology numerical simulations represent an alternative method to 
study dynamics of complex systems and to predict their behavior in non-equilibrium conditions. 
In the problems of pattern formation one usually deals with reaction-diffusion systems which play an important 
role in the study of the generic spatiotemporal behavior of nonequilibrium systems. These models contain main
contributions related to both local dynamics (chemical reactions of a birth-and-death type) and mass transport.
This approach gives a possibility to control dynamics of self-organization of an ensemble of adatoms, morphology 
of the growing surface, shape of structures and linear size of adsorbate (vacancy) islands. 
Previously, by using numerical simulations, we have shown that the morphology of the surface and the linear size of 
adsorbate structures in multi-layer plasma-condensate systems can be 
controlled by the strength of the electrical field near substrate \cite{NRL17}. 

It is known, that fluctuations can play a crucial role in problems of pattern formation and self-organization, in general.
Fluctuation-induced effects were studied in different classes of problems \cite{VanKampen,GarciaOjalvo}. 
By considering adsorbate clusters formation during deposition, fluctuations usually assumed to be 
negligibly small. Nevertheless, it was shown previously, that fluctuations of the lateral adsorbate flux 
at epitaxial growth result in the formation of patterns with a needle-like structure \cite{PhysScr11}. 
During adsorption in stochastic model of gas-condensate systems the internal noise satisfying the 
fluctuation-dissipation relation governs phase transitions between the dense and the diluted phases and controls 
type of surface patterns \cite{PRE12}.

In this work we will derive the stochastic one-layer model to describe pattern formation 
on intermediate layer of multi-layer plasma-condensate systems by taking into account anisotropy in transference of adatoms 
between neighbor layers, induced by the electric field near substrate and fluctuations of the lateral flux of adsorbate. 
By using stability analysis we will show that this noise leads to homogenization of adsorbate distribution 
on the intermediate layer. By using numerical simulations we will discuss an influence of the introduced fluctuation's 
intensity onto dynamics of pattern formation, morphology of the growing surface and statistical properties of 
the surface structures.

The work is organized in the following manner. In the next section we will derive the stochastic one-layer model. 
Section 3 contains results of the stability analysis. In Section 4 we perform numerical simulations. We conclude 
in the last Section.

\section{Model}

To describe an evolution of the adsorbate concentration on the substrate at multi-layer condensation 
we will monitor the coverage field $x_n({\bf r},t)\in[0,1]$ on each $n$-th layer, defined as the ratio 
of the adsorbed particles (atoms, ions) and possible free sites in a unit cell on the surface; 
${\bf r}=\{x,y\}$ is the space coordinate and $t$ is the adsorption time. Hence, the spatio-temporal 
change in the adsorbate concentration can be described by the model of the reaction-diffusion 
system in the standard form:
\begin{equation}
\partial x_n({\bf r},t)=R(x_n)-\nabla\cdot{\bf J}_n,
\label{eq1}
\end{equation}  
where $R(x_n)$ is the reaction term, describing ``born-death'' processes and ${\bf J}_n$ is the adsorbate flux, 
related to the mass transport; index $n$ defines the layer number of the multi-layer system. 

According to the 
standard approach for description of the condensation process the reaction term $R(x_n)$ includes 
adsorption, desorption and transference reactions between neighbor layers.  Adsorption processes 
are described by the term $k_ax_{n-1}(1-x_n)(1-x_{n+1})$ with adsorption coefficient 
$k_a=\varpi p\exp(-E_a/T)$ defined through the pressure $p$, activation energy for adsorption $E_a$, 
temperature $T$, measured here in energetic units, and frequency factor $\varpi$. 
These processes require free $(1-x_n)$ sites on the current layer, 
free $(1-x_{n+1})$ sites on the next layer and non-zero concentration $x_{n-1}$ on the precursor layer. 
Adsorbed particles can desorb from the layer with the rate 
$k_d=k_d^0\exp(-U_n/T)$ defined by the the desorption rate for noninteracting particles 
$k_d^0=\varpi \exp(-E_d/T)$, where $E_d$ is the desorption activation energy, and the interaction potential 
$U_n({\bf r})$. The life time scale of adatoms $\tau_d$ relates to the $k_d^0$ as $\tau_d=[k_d^0]^{-1}$. 
By considering desorption mediated by the precursor layer the desorption term attains the form 
$-k_dx_nx_{n-1}(1-x_{n+1})$. Adatoms can move between neighbor layers with the probabilities of motion 
from top towards bottom layers $\omega_\downarrow$ and back $\omega_\uparrow$. In the case 
$\omega_\downarrow=\omega_\uparrow$ one gets the standard vertical diffusion \cite{CWM2002}. 
In the general case, these probabilities can be different. For example, in low-vacuum gas-condensate systems 
the pressure of the gaseous phase induces the anisotropy in vertical motion with preferential motion from top 
layers towards bottom ones, leading to $\omega_\uparrow/\omega_\downarrow<1$ \cite{SS15}. 
In plasma-condensate systems the electric field near substrate induces opposite motion with
$\omega_\uparrow/\omega_\downarrow>1$ \cite{NRL17}. In the last case one has 
$\omega_\uparrow=\omega_\downarrow+\omega_E$, where the anisotropy strength 
$\omega_E=|{\bf E}|Ze/T$ is proportional to the strength of the electric field near substrate $|{\bf E}|$; 
$Z$ is the coordination number and $e$ is the electron charge. By assuming that the transference is possible onto 
free sites only, for the transference reactions we get 
$\omega_\downarrow(x_{n-1}+x_{n+1}-2x_n)+\omega_E[x_{n-1}(1-x_n)-x_n(1-x_{n+1})]$, 
where the first term corresponds to the standard vertical diffusion and the second one relates to the 
electric field induced anisotropy.  
   
The lateral diffusion flux on the $n$-th layer ${\bf J}_n$ includes free later diffusion of adsorbate $-D_\leftrightarrow\nabla x_n$ 
with lateral diffusion coefficient $D_\leftrightarrow$ and diffusion caused by the adsorbate interactions with potential $U_n(r)$ 
in the form $-D_\leftrightarrow/T\mu(x_n)\nabla U_n$, where $\mu(x_n)=x_n(1-x_n)$ indicates that this diffusion is possible 
on free sites, only. Following Refs.\cite{Wolgraef2003,Wolgraef2004,PRE12,PhysScr12,SS14,SS15,NRL17,CWM2002}
the interaction potential $U_n(r)$ can be defined in the framework of self-consistence approximation by 
attractive substratum mediated potential $u(r)$ among adsorbed particles separated by a distance $r$ in a layer, as 
$U_n(r)=-x_{n-1}\int u(r-r')x_n(r'){\rm d}r'$. Next, by using Gaussian profile 
$u(r)=2\epsilon(4\pi r_{0}^2)^{-1/2}\exp\left(-r^2/4r_{0}^2\right)$ with  
interaction energy $\epsilon$ and interaction radius $r_0$ and assuming that the adsorbate concentration varies slow 
within the interaction radius we expand the integral $\int u(r-r')x_n(r'){\rm d}r'$ and retain three non-vanishing terms. 
Finally, for the interaction potential $U(r)$ we get: 
\begin{equation}
U(r)\simeq-\epsilon x_{n-1}[x_n+(1+r_0^2\nabla^2)^2x_n].
\label{eq2}
\end{equation}

In the recent work \cite{EPJB18} we have derived the adsorbate concentration on both $(n+1)$-th and 
$(n-1)$-th layers through one on the $n$-th layer, by considering adsorbate concentration 
on the any $n$-th layer as $x_n=S_n/S_0$, where $S_0$ and $S_n$ are squares of the substrate and 
and $n$-th layer. By taking into account that the adsorbate 
concentration decreases with the layer number according to the principle of the surface energy minimization 
the linear size $R_n$ of the multi-layer pyramidal-like adsorbate structure decreases with the layer number $n$ 
as $R_n=R_0-nd$, where the width of the terrace $d$ defines the total number $N$ of layers in multi-layer 
adsorbate structure. In such a case one gets $x_n=[1-n(d/R_0)]$ with $R_0^2\propto S_0$
and $x_{n\pm1}=[1-(n\pm1)(d/R_0)]$. After the simple algebra one finds: 
\begin{equation}
x_{n\pm1}=\left(\sqrt{x_n}\mp\beta_0/2\right)^2,
\label{eq3}
\end{equation}
where $\beta_0=2d/R_0<1$. Finally, by using relations (\ref{eq3}) the reaction part $R(x=x_n)$ attains the form:
\begin{equation}
\begin{split}
 R(x)&=\alpha(1-x)\nu(x)-x\nu(x)e^{-2\varepsilon x(\sqrt{x}+\frac{1}{2}\beta_0)^2}\\ 
 &+u\beta_0\sqrt{x}(1-2x)+\frac{1}{4}\beta_0^2(u+2D_\updownarrow),
\end{split}
\label{eq4}
\end{equation}
where the dimensionless constants 
$\varepsilon\equiv\epsilon/T$, $\alpha\equiv k_{a}/k_d^0$, $u=\omega_E/k_d^0$, 
$D_\updownarrow\equiv \omega_\downarrow/k_d^0$ are used and 
$\nu(x)=(\sqrt{x}+1/2\beta_0)^2\left[1-(\sqrt{x}-1/2\beta_0)^2\right]$ 
is introduced. The total lateral adsorbate flux ${\bf J}$ reads:
\begin{equation}
{\bf J}=-D_\leftrightarrow\left[ \nabla x-\varepsilon\gamma(x)\nabla\left\{x+(1+r_0^2\nabla^2)^2x\right\}\right]. 
\label{eq5}
\end{equation} 
It can be separated into the two terms: ${\bf J}_1=-D_\leftrightarrow(1-\varepsilon\gamma(x))\nabla x$ and 
${\bf J}_2=D_\leftrightarrow\varepsilon\gamma(x)\mathcal{L}_{SH}x$, where the notation for the 
Swift-Hohenberg operator $\mathcal{L}_{SH}=(1+r_0^2\nabla^2)^2$ is used \cite{LSH}. 
Formally, ${\bf J}_2$ describes an influence of the micro-level (interacting adatoms) onto the 
pattern formation on the meso-level, where the description of the system dynamics is provided 
by using concentration of the adsorbate as the man variable. Hence, in the general case the flux ${\bf J}_2$ 
has both regular and stochastic parts: 
\begin{equation}
{\bf J}_2=D_\leftrightarrow\varepsilon\gamma(x)\mathcal{L}_{SH}x+\zeta(x;{\bf r},t),
\label{eq6}
\end{equation}
where for the stochastic term $\zeta(x;{\bf r},t)$ we assume Gaussian properties: 
$\langle\zeta(x;{\bf r},t)\rangle=0$, 
$\langle\zeta(x;{\bf r},t)\zeta(x;{\bf r}',t')\rangle=2\Sigma \mathcal{D}(x)\delta({\bf r}-{\bf r}')\delta(t-t')$.
Here $\mathcal{D}(x)=D_\leftrightarrow\varepsilon\gamma(x)$ and $\Sigma$ is the noise intensity.
Thus, by combining Eqs.(\ref{eq4})-(\ref{eq6}), and introducing the diffusion length
$L_d\equiv\sqrt{D_\leftrightarrow/k_d^0}$ the evolution equation fro the adsorbate concentration 
on the intermediate layer of the multi-layer plasma-condensate system in the Stratonovich interpretation reads:
\begin{equation}
\begin{split}
\frac{\partial x}{\partial t}&=R(x)-L_d^2\nabla\cdot{\bf J}\\ 
&-\frac{\Sigma}{2}\nabla\cdot\left(\nabla \frac{{\rm d}\mathcal{D}}{{\rm d}x}\right)
+\nabla\sqrt{\mathcal{D}(x)}\xi({\bf r},t)
\end{split}
\label{eq7}
\end{equation}  
with $\langle\xi({\bf r},t)\rangle=0$ and $\langle\xi({\bf r},t)\xi({\bf r}',t')\rangle=2\Sigma\delta({\bf r}-{\bf r}')\delta(t-t')$. 
The main goal of this work is to perform a detailed study of an influence of the introduced fluctuations onto 
dynamics of pattern formation in the studied system and statistical properties of the surface morphology. 

\section{Results and discussions}

\subsection{Linear stability analysis}

\begin{figure*}
a)\includegraphics[width=0.3\textwidth]{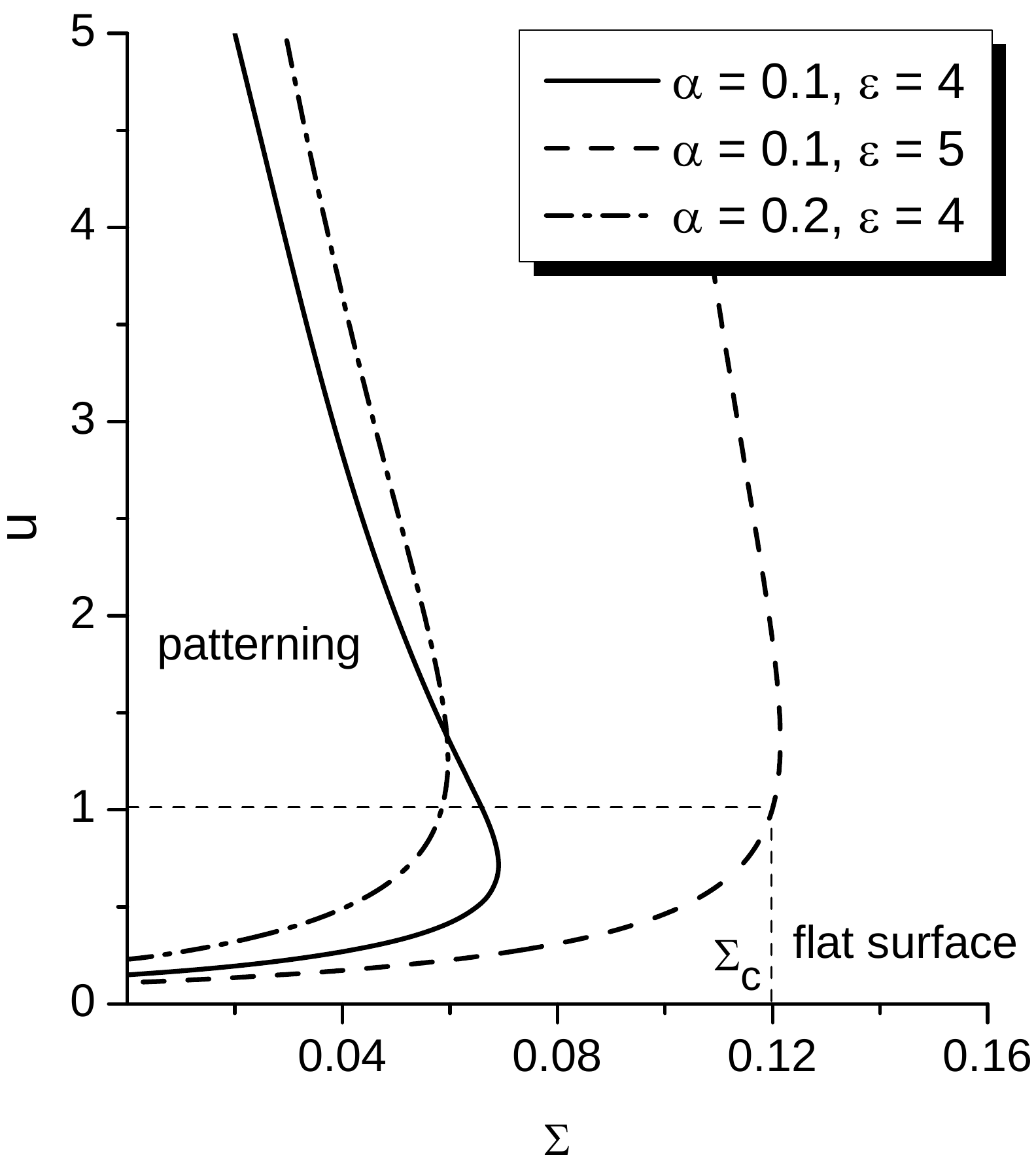}
b)\includegraphics[width=0.3\textwidth]{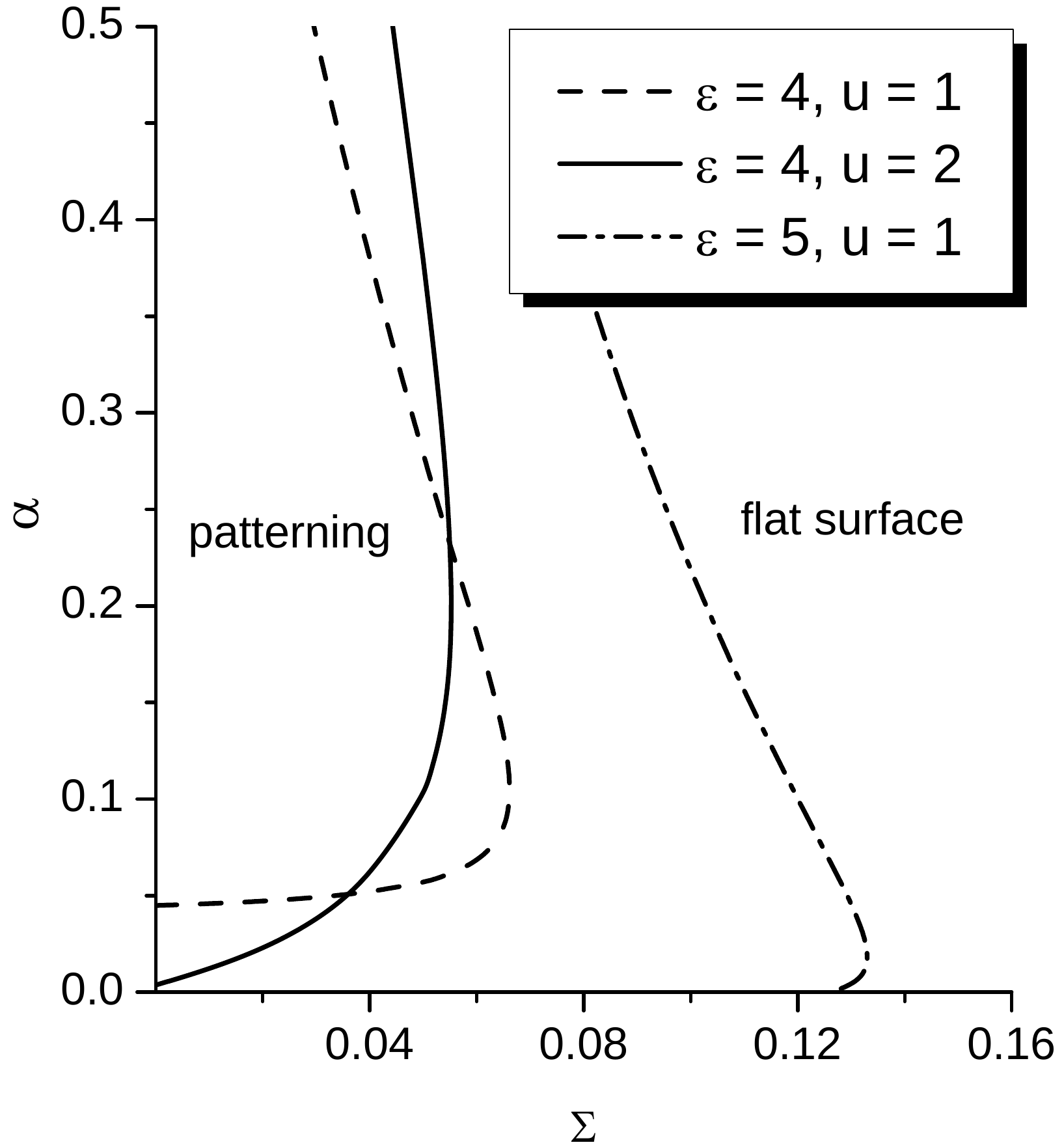}
c)\includegraphics[width=0.3\textwidth]{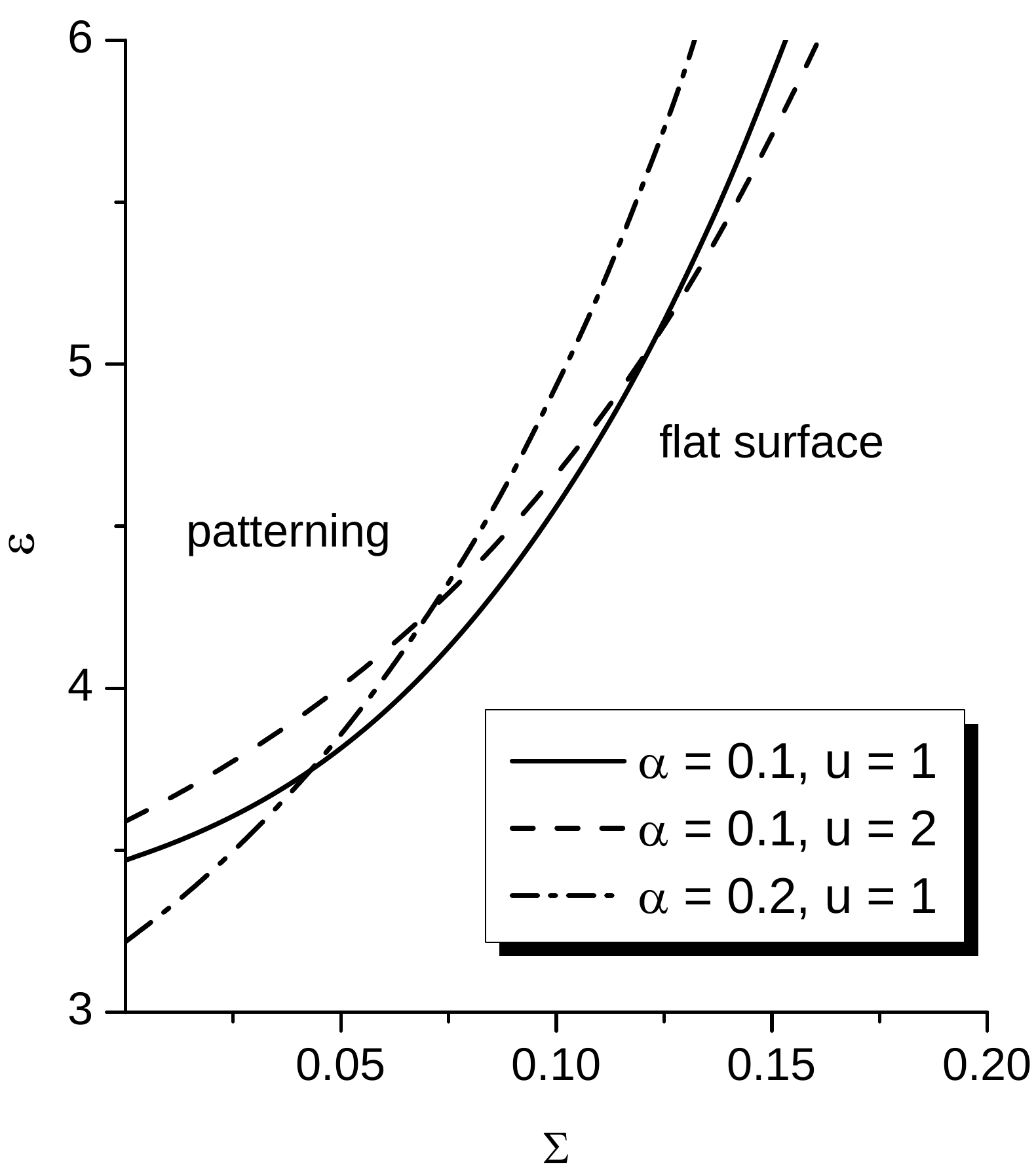}
\caption{Stability diagrams: in domain ``patterning'' separated surface structures are realized; in domain 
``flat surface'' adsorbate homogeneously covers the layer. }
\label{fig1}
\end{figure*}

The linear stability analysis allows one to define domains of the main system parameters, 
when the pattern formation is possible. For the actual problem it was shown previously, that 
if the anisotropy strength of the vertical diffusion is infinitely small ($u\to0$), that corresponds 
to the weak electric field near substrate, then adsorbate homogeneously covers the intermediate layer 
and multi-layer layers adsorbate islands could not be organized \cite{EPJB18}. In this study 
we will focus onto an influence of the fluctuations intensity $\Sigma$ onto stability of the 
stationary homogeneous states $x_{0}$ to inhomogeneous perturbations. As it follows from Eq.(\ref{eq7}) 
the introduced fluctuations do not influence $x_{st}$, which can be found from the relation $R(x_{0})=0$.
According to the standard linear stability analysis one assumes the deviation of the adsorbate concentration 
$x$ from the stationary value $x_{0}$ in the form $\delta x=x-x_{0}\propto e^{\lambda(k)t}e^{ikr}$, 
where the stability exponent $\lambda(k)$ depends on the wave number $k$. Expanding the reaction term 
$R(x)$ in the vicinity of $x_{0}$ one gets:
\begin{equation}
\begin{split}
&\lambda(\kappa)={\rm d}_xR(x)|_{x=x_{st}}\\
  &-\kappa^2\left[1-2\varepsilon\gamma(x_{0})(1-\rho^2\kappa^2)-\frac{\Sigma}{2}\varepsilon\gamma(x_{0})\right],
\label{eq8}
\end{split}
\end{equation}
where notations $\kappa\equiv kL_d$, $\rho\equiv r_0/L_d$ are used and the limit $\rho^4\to0$ is considered 
\footnote{Estimation for the interaction radius and diffusion length gives: $r_0\sim1$nm, $L_d\sim1\mu$m 
\cite{Mikhailov1,Mikhailov2}.}. In the case $\lambda(\kappa)<0\ \forall\ \kappa$ no stable separated structures 
can be formed. If $\lambda(\kappa)>0$ at $\kappa\in[0,\kappa_c]$ then the phase separation scenario is realized.
In the case $\lambda(\kappa)>0$ at $\kappa\in[\kappa_1,\kappa_2]$ separated structures will be stable with the mean 
distance between them $\kappa_0^{-1}$ where $\lambda(\kappa_0)$ has maximal value. By analyzing stability 
exponent $\lambda(\kappa)$ we have calculated the stability diagrams, shown in Fig.\ref{fig1}.
\begin{figure}
\includegraphics[width=0.9\columnwidth]{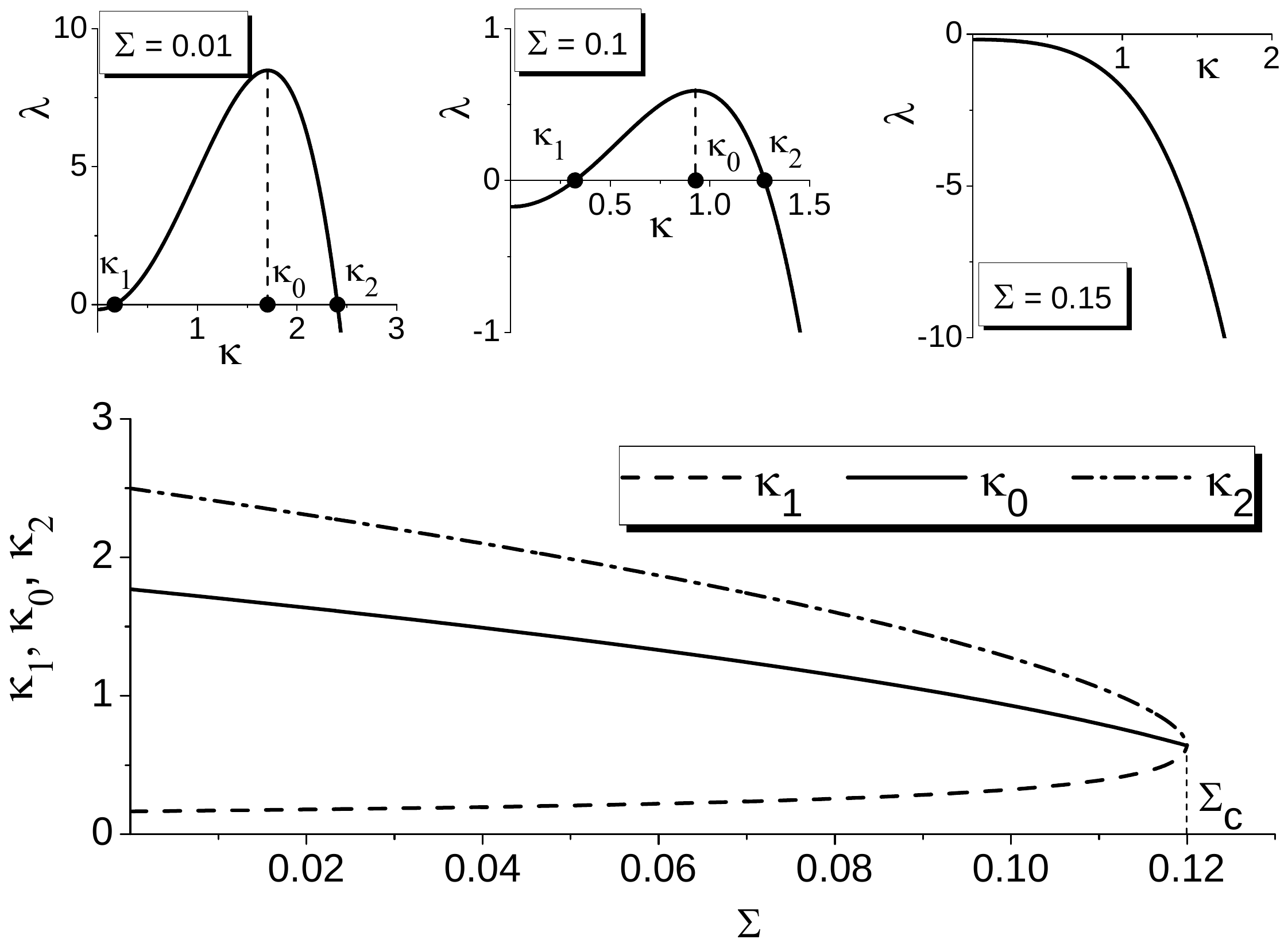}
\caption{Dependencies of the stability exponent $\lambda$ on the reduced wave number $\kappa$ 
at $\varepsilon=5$, $u=1$, $\alpha=0.1$ and different values of the noise intensity  $\Sigma$ 
(in the top); and dependencies of the wave numbers $\kappa_1$, $\kappa_0$, $\kappa_2$ on 
$\Sigma$ (bottom panel).}
\label{fig2}
\end{figure}

From Fig.\ref{fig1}a it follows, that an increase in the noise intensity leads to the transition from the domain 
of pattern formation ``patterning'' towards domain of the homogeneous distribution of the adsorbate on the 
layer ``flat surface'' at critical value $\Sigma_c$. By analyzing Figs.\ref{fig1}a,b,c one finds that 
at fixed noise intensity a growth in the anisotropy strength $u$ 
or adsorption coefficient $\alpha$ acts in the same manner; an increase in the interaction strength extends 
the domain of stable surface structures formation (detailed description about an influence of the main system 
parameters onto stability diagram in deterministic plasma-condensate system was done in Ref.\cite{EPJB18}). 
Dependencies of the stability exponent $\lambda(\kappa)$ at different values of the noise intensity and 
reduced wave-numbers $\kappa_1$, $\kappa_0$ and $\kappa_2$ on fluctuation's intensity 
$\Sigma$ at other fixed parameters are shown in Fig.\ref{fig2}.
From the top panel it follows, that with the noise intensity growth the dependence $\lambda(\kappa)$ moves 
down towards negative values, meaning transition to the homogeneous configuration. The falling-down dependence 
of the period of spatial modulations $\kappa_0$ indicates an increase in the mean distance between 
separated surface structures. It is important, that at $\Sigma=\Sigma_c$ one has $\kappa_0\ne0$. It means, 
that in the limit $\Sigma\to\Sigma_c-0$ the surface of the layer is characterized by the finite number of the 
surface structures with the fixed mean distance between them. 

The provided stability analysis could not give any information about the type and size of the surface structures 
and dynamics of the adsorbate distribution. It only predicts the possibility of pattern formation in the 
stationary limit.  

\subsection{Numerical simulations}

Next we will use numerical simulations to perform detailed study an influence of the fluctuation's intensity 
$\Sigma$ onto dynamics of pattern formation and possibility of islands size control in stochastic 
plasma-condensate systems. To that end we fix $\beta=0.1$, $u=1.0$, $\epsilon=5.0$, $\alpha=0.1$ and 
solve numerically the Langevin equation (\ref{eq7}) on two-dimensional hexagonal grid with linear size 
$L=256\Delta x$ with periodic boundary conditions. 
The time step was $\Delta t=10^{-3}$, the spatial integration step was $\Delta x=0.5$. As initial conditions we use 
Gaussian distribution with $\langle x({\bf r},0)\rangle=\langle (\delta x({\bf r},0))^2\rangle=10^{-2}$. 
By using in the computational scheme $\rho=0.025$ the total size of the system is $L=12.8L_d$.  

\begin{figure*}
a)\includegraphics[width=0.8\textwidth]{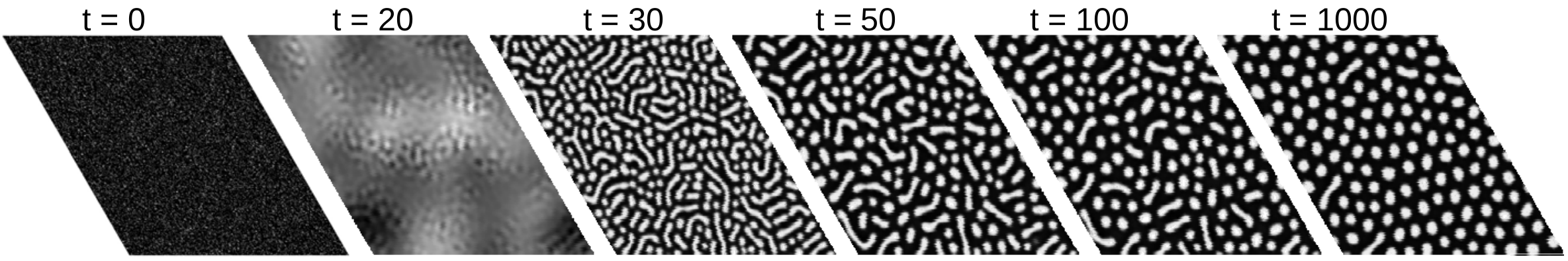}\\
b)\includegraphics[width=0.8\columnwidth]{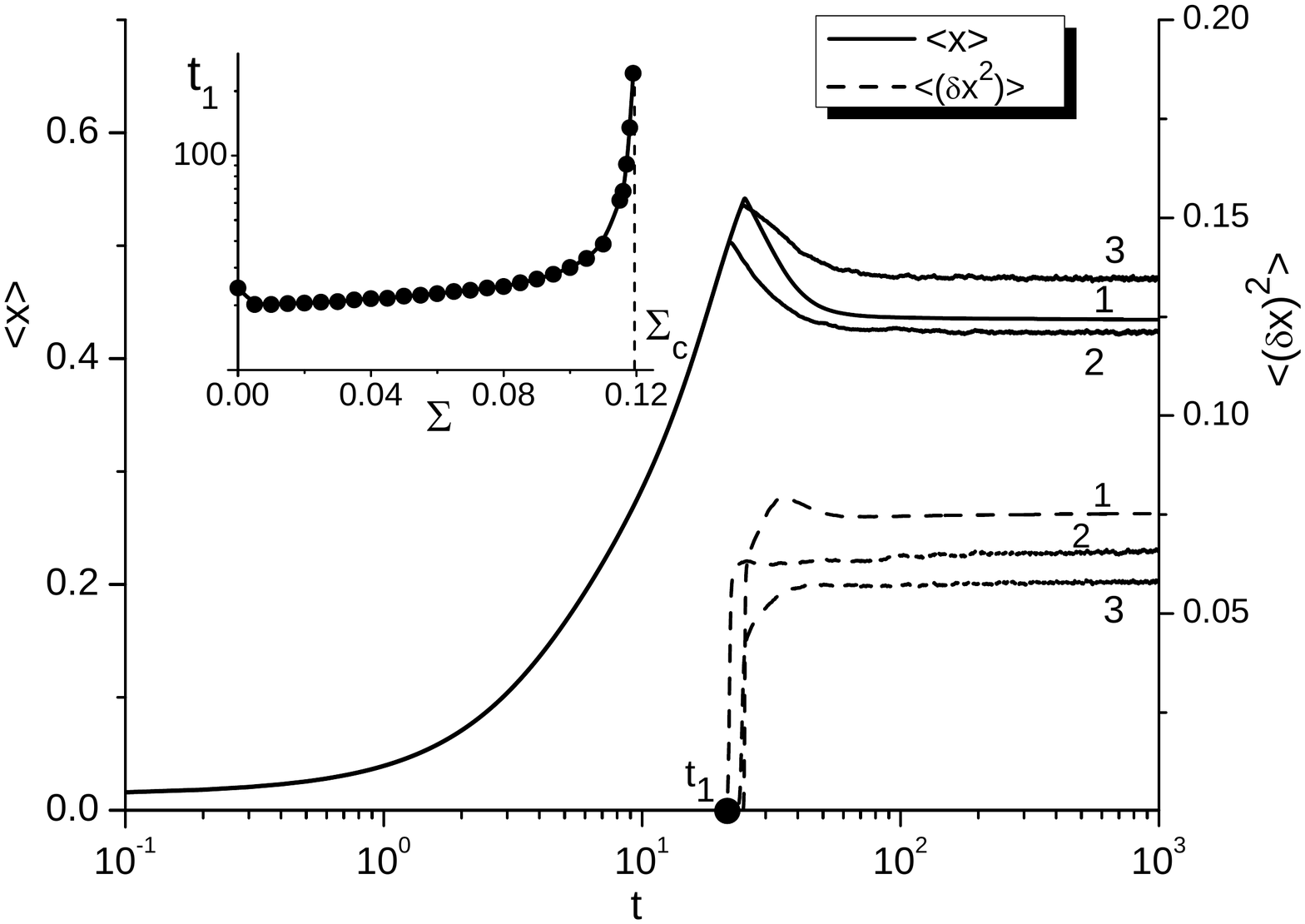}
c)\includegraphics[width=0.8\columnwidth]{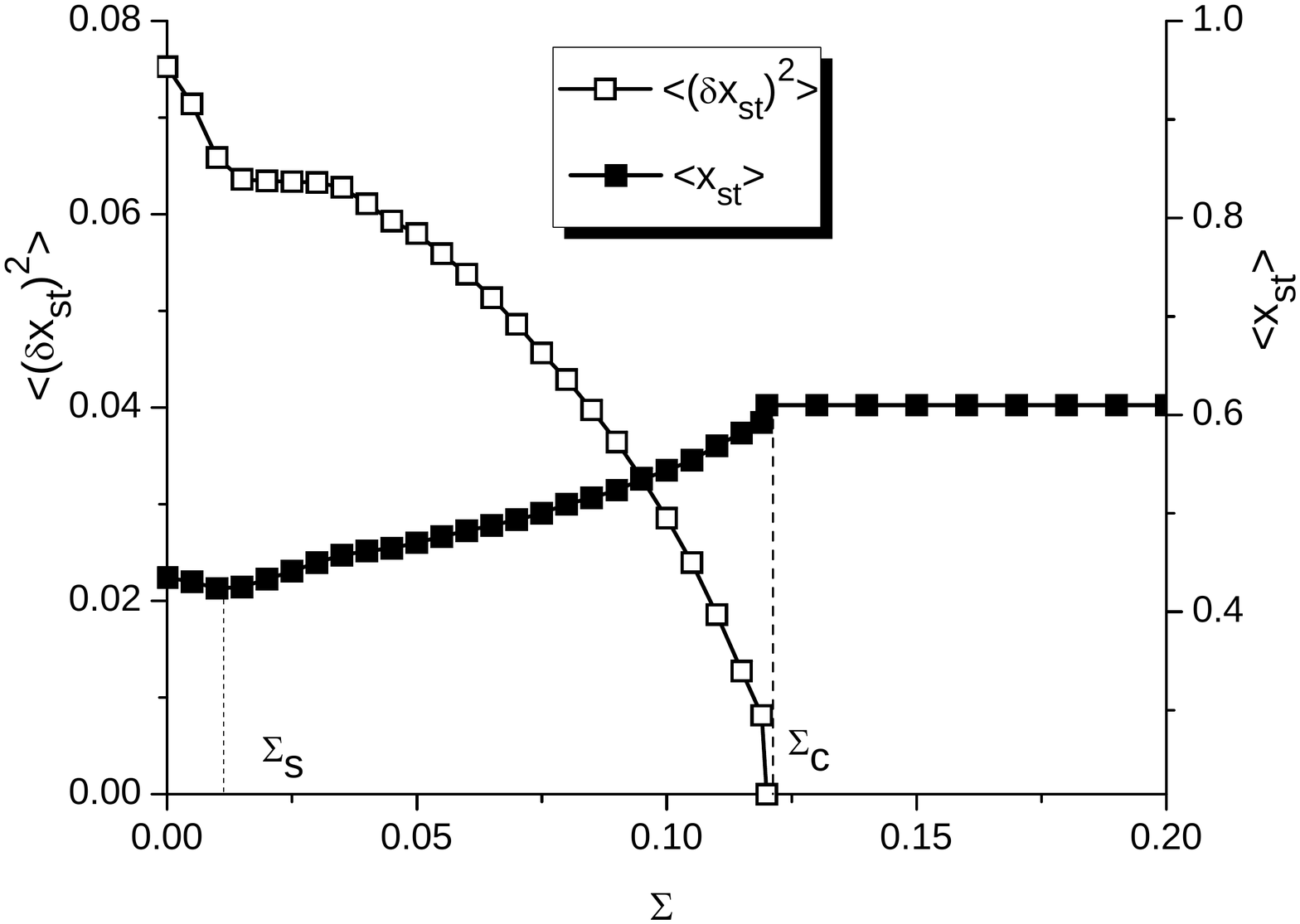}
\caption{(a) Snapshots of the system evolution at $\Sigma=0.05$. 
(b) Evolution of the mean adsorbate concentration $\langle x \rangle$ and 
order parameter $\langle(\delta x)^2\rangle$ at: 1) $\Sigma=0$, 2) $\Sigma=0.01$, 3) $\Sigma=0.1$. 
Insertion shows dependence of the time instant $t_1$, when the order parameter starts to grow, on 
the noise intensity $\Sigma$. 
(c) Dependencies of the mean stationary values of adsorbate concentration $\langle x_{st}\rangle$ 
and order parameter $\langle(\delta x_{st})^2\rangle$ on the noise intensity $\Sigma$.}
\label{fig3}
\end{figure*}

In Fig.\ref{fig3}a we show snapshots of the system evolution at $\Sigma=0.05$. It follows, that during exposing 
the initially Gaussian surface (picture at $t=0$) starts to structuring (picture at $t=20$) and small adsorbate 
islands of different form emerge (shown by white color in picture at $t=30$). These islands grow and interact 
forming during system evolution pattern with separated adsorbate islands with approximately the same 
shape and linear size (see pictures at $t=50$, 100, 1000 in Fig.\ref{fig3}a). In order to perform 
a detailed study of the system dynamics and characterize an influence of the fluctuation's intensity $\Sigma$ 
we consider dynamics of first two statistical moments, namely, mean adsorbate concentration on a layer 
$\langle x \rangle$ and dispersion $\langle (\delta x)^2\rangle$, shown in Fig.\ref{fig3}b at different 
values of the noise intensity $\Sigma$. It follows that at initial stages of the system evolution 
fluctuations have no any influence on the mean concentration $\langle x\rangle$, which grows with time. 
At the time instant $t=t_1$ the adsorbate supersaturation becomes enough and interaction processes 
become to play a dominant role in the system dynamics, leading to formation of the separated adsorbate islands 
(see picture at $t=20$ in Fig.\ref{fig3}a). Here $\langle x\rangle$ takes maximal value and with further 
exposing decreases towards the non-zero stationary value $\langle x_{st}\rangle$, which does not change 
at the large time scales and depends on the $\Sigma$. A decreasing of the mean adsorbate concentration 
relates to the interaction of adsorbate islands: small ones with the linear size less than the critical one disappear 
whereas large islands grow until the linear size becomes the stationary value, depending on the system parameters. 
Dispersion $\langle (\delta x)^2\rangle$, which plays a role 
of the order parameter in problems of pattern formation and phase separation, takes zero value until $t<t_1$, 
meaning homogeneous distribution of adsorbate on the layer. At $t=t_1$ the quantity $\langle (\delta x)^2\rangle$ 
starts to growth, indicating ordering processes on the layer. At large time scales  $\langle (\delta x)^2\rangle$  
attains stationary non-zero value $\langle (\delta x_st)^2\rangle$, depending on the fluctuation's intensity 
$\Sigma$, meaning formation of stable patterns on the layer. 
The value $t_t$, when ordering processes start, depends on the noise intensity $\Sigma$, that is 
shown in the insertion in Fig.\ref{fig3}b. It follows, that an introducing weak fluctuations with small 
intensity $\Sigma=0.01$ leads to accelerating of the patterning (a decrease in $t_1$) with a little bit lower value 
of the mean adsorbate concentration, comparing to the deterministic case with $\Sigma=0$ (see insertion and 
solid curves 2 and 1).  A further increase in the noise intensity results in delaying dynamics of pattern formation 
due to increasing dependence $t_1(\Sigma)$ (see insertion in Fig.\ref{fig3}b);  $\langle x_{st}\rangle$ grows 
(see solid curve 3). From the other hand $\langle (\delta x_st)^2\rangle$ decreases with $\Sigma$ (see dash 
curves in Fig.\ref{fig3}b).
\begin{figure*}
\begin{minipage}{0.49\textwidth}
a)\includegraphics[width=0.9\textwidth]{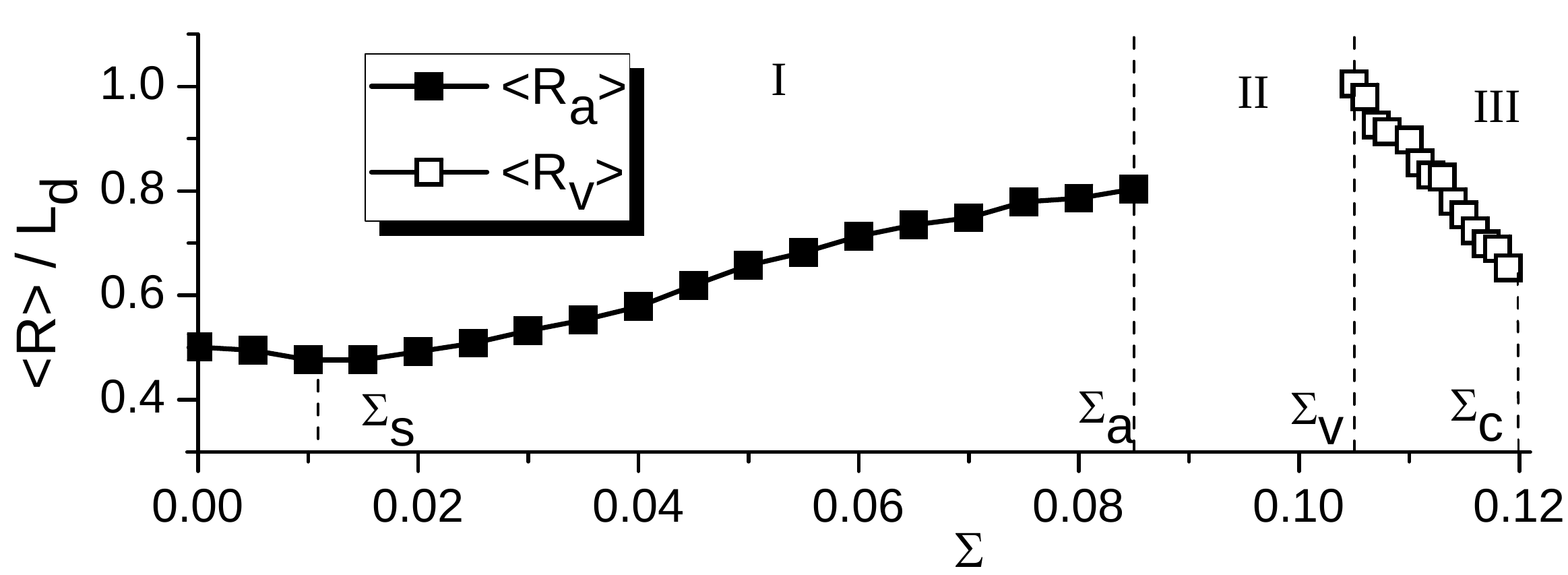}\\
b)\includegraphics[width=0.9\textwidth]{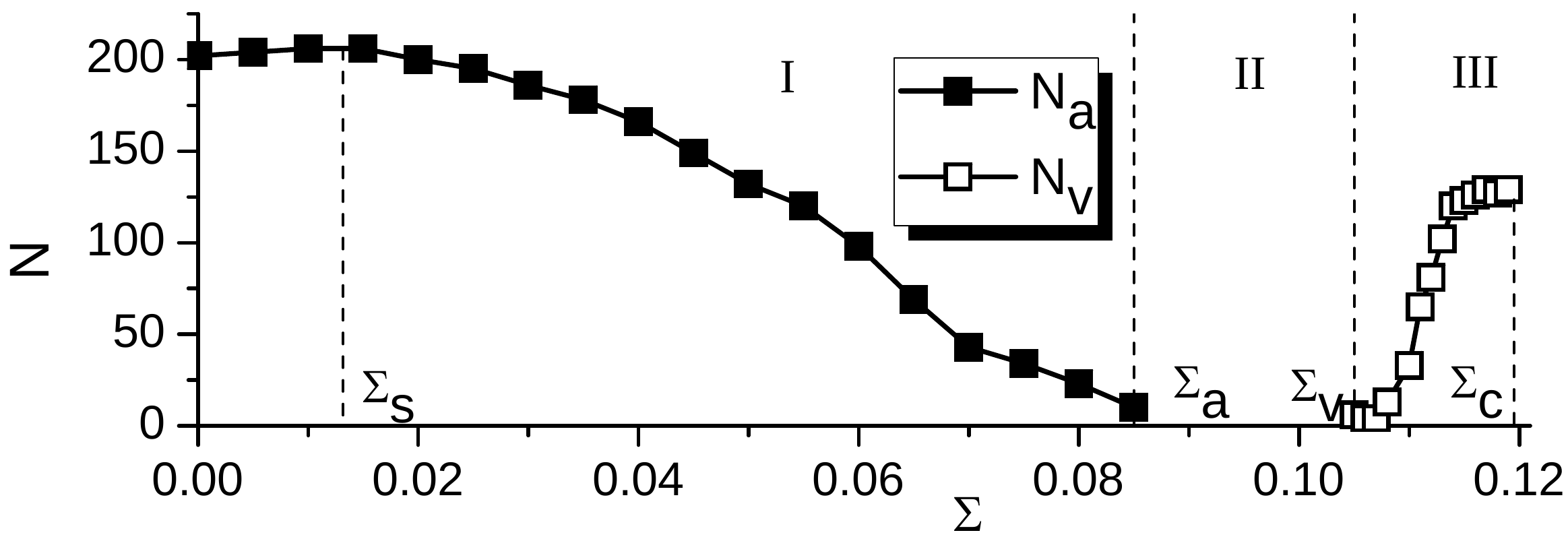}
\end{minipage}
\begin{minipage}{0.49\textwidth}
c)\includegraphics[width=0.9\textwidth]{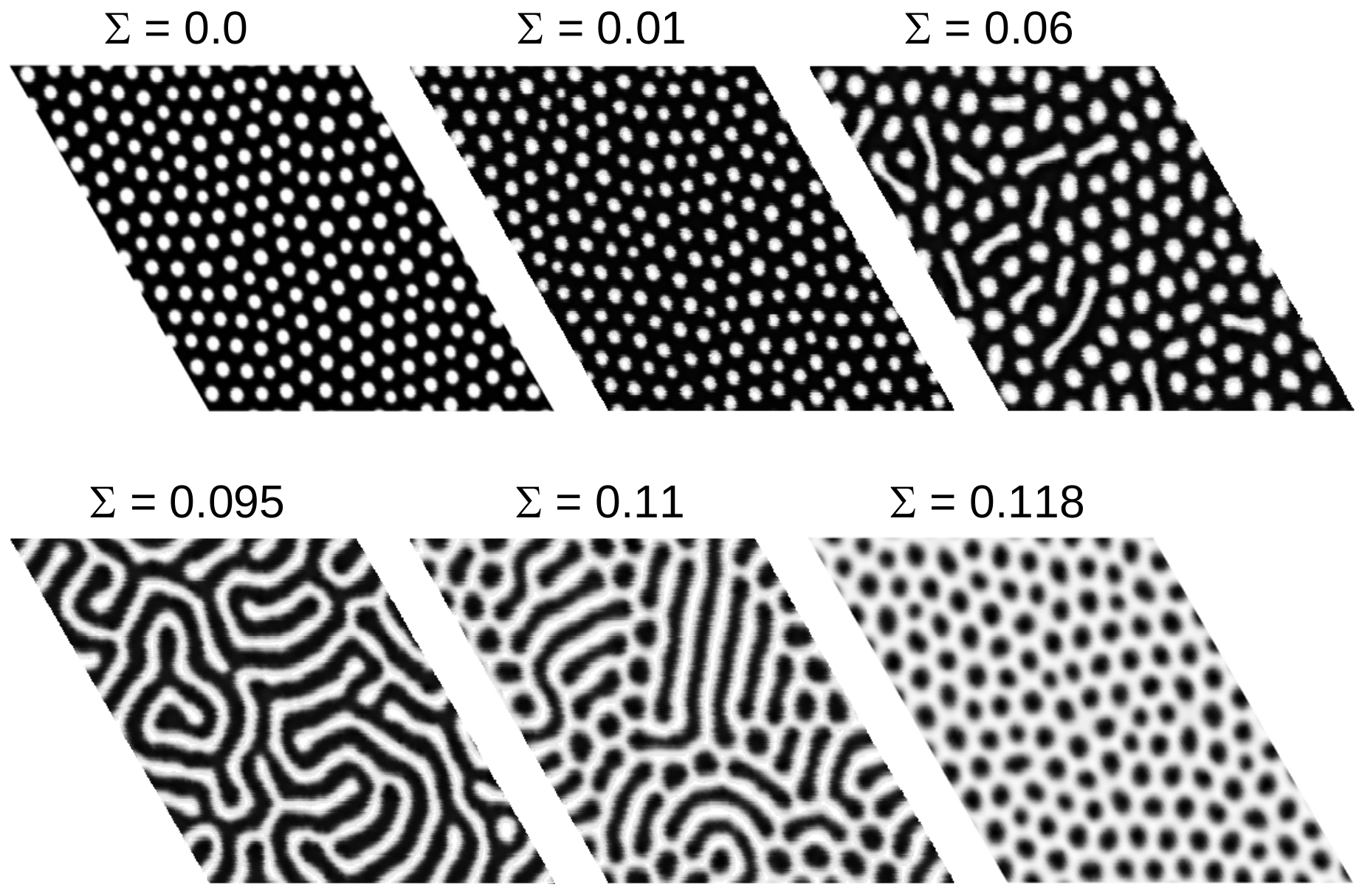}
\end{minipage}
\caption{Dependencies of (a) the mean linear size (radius) $\langle R \rangle$ and 
(b) the number of spherical structures of adsorbate (filled squares) and vacancy islands (empty squares) 
on the noise intensity $\Sigma$. (c) Stationary snapshots at different $\Sigma$.}
\label{fig4}
\end{figure*}

In Fig.\ref{fig3}c we plot dependencies of both the stationary adsorbate concentration $\langle x_{st}\rangle$ 
and stationary order parameter $\langle(\delta x_{st})^2\rangle$ on fluctuation's intensity $\Sigma$. It is 
seen, that with the $\Sigma$ growth the quantity $\langle x_{st}\rangle$ decreases, attains minimal value 
at $\Sigma=\Sigma_s$ and then increases until $\Sigma=\Sigma_c$ (see filled squares in Fig.\ref{fig3}c). 
The quantity  $\langle(\delta x_{st})^2\rangle$ decreases with $\Sigma$ and attains zero values at 
$\Sigma=\Sigma_c$ (empty squares in Fig.\ref{fig3}c). It means, that if $\Sigma>\Sigma_c$ then 
no stationary separated structures can be formed on the layer; here noise intensity does not influence the mean 
adsorbate concentration (filled squares in Fig.\ref{fig3}c at $\Sigma>\Sigma_c$). 

In order to analyze an influence of the fluctuation's intensity $\Sigma$ on a type of the surface structures, 
their number and linear size we use the percolating clusters method and define the square of each cluster on the 
half-height of the layer and their number. Next, by assuming that surface structures are of spherical form 
we calculate the linear size (radius) in units of diffusion length $L_d$. In this procedure we do not take into 
account large elongated clusters. Obtained results, as dependencies of the mean linear size of spherical  
islands $\langle R\rangle$ and their number $N$ on the fluctuation's intensity $\Sigma$ are shown in 
Figs.\ref{fig4}a,b, respectively. Typical stationary snapshots at different values of $\Sigma$ are shown in Fig.\ref{fig4}c.
From Figs.\ref{fig4}a,b it follows that on both dependencies $\langle R\rangle(\Sigma)$ and $N(\Sigma)$ 
one can issue three domains, indicated as I, II and III at $\Sigma\in[0,\Sigma_c)$ (at $\Sigma>\Sigma_c$ 
no structures can be formed).  In the domain I at small values of the noise intensity one has separated 
sphere-shaped adsorbate islands on the layer (see snapshots at $\Sigma=0$ and 0.01 in Fig.\ref{fig4}c). 
Here at $\Sigma<\Sigma_s$ fluctuations provide a decrease in the linear size of adsorbate islands and an 
increase in their number. At $\Sigma_s<\Sigma<\Sigma_a$ noise action leads to a decrease in the number 
of islands and growth of their size. Moreover an increase in the noise intensity in domain I provides formation 
of elongated adsorbate islands (compare snapshots at $\Sigma=0.01$ and $\Sigma=0.06$ in Fig.\ref{fig4}c). 
In the domain II, when $\Sigma_a<\Sigma<\Sigma_v$, one has surface pattern of percolating adsorbate island, 
like in phase decomposition scenario (see snapshot at $\Sigma=0.095$ in Fig.\ref{fig4}c). With further $\Sigma$ 
growth, in domain III, separated vacancy islands start to organize inside adsorbate matrix (see snapshot at 
$\Sigma=0.11$ in Fig.\ref{fig4}c). Here an increase in noise intensity leads to formation of spherical-shaped 
vacancy islands and their linear size decreases (compare snapshots at $\Sigma=0.11$ and $\Sigma=0.118$ 
in Fig.\ref{fig4}c). Large fluctuations with $\Sigma>\Sigma_c$ makes pattern formation impossible and adsorbate 
homogeneously covers the layer. Hence, By varying the fluctuation's intensity one can control morphology of 
the growing surface and linear size of adsorbate or vacancy islands. Analogous effects of transition from separated 
adsorbate islands towards separated vacancy islands in adsorbate matrix were found by varying the pressure 
of the gaseous phase in on-layer model of gas-condensate system \cite{SS14}.    
It should be noted, that the critical value $\Sigma_c\simeq0.12$ obtained in the 
framework of linear stability analysis (see Figs.\ref{fig1}a and Fig.\ref{fig2}) corresponds well to one, 
coming from numerical simulations at fixed other system parameters. 
    
\section*{Acknowledgments}

Support of this research by the Ministry of Education and Science of Ukraine, 
project No. 0117U003927, is gratefully acknowledged.

\section{Conclusions}

In this article we derived the stochastic model of plasma-condensate systems by introducing fluctuations 
of the adsorbate flux composed of ordinary diffusion flux and the flux related to interaction of adatoms. 
The proposed model is used to study pattern formation on an intermediate layer of multi-layer 
plasma-condensate system with anisotropy in transference of adatoms between neighbor layers induced 
by the electrical field presence nearby substrate. In the framework of the stability analysis and by numerical 
simulations we have shown that an increase in the fluctuation's intensity leads to homogenization of the 
adsorbate distribution on the layer. It is shown that in the case of weak fluctuations an increase in the noise 
intensity leads to an acceleration in pattern formation and to a decrease in the linear size of separated 
spherical-shaped adsorbate islands. Further increase in the noise intensity results in delaying dynamics 
of patterning, an increase in a linear size of adsorbate islands, formation elongated adsorbate islands, 
transition from the configuration of separated adsorbate islands on a layer towards separated vacancy 
islands inside adsorbate matrix and homogenization of the adsorbate distribution on a layer.



\end{document}